\documentclass[sn-mathphys,Numbered]{sn-jnl}

\usepackage{graphicx}%
\usepackage{multirow}%
\usepackage{amsmath,amssymb,amsfonts}%
\usepackage{amsthm}%
\usepackage{mathrsfs}%
\usepackage[title]{appendix}%
\usepackage{xcolor}%
\usepackage{textcomp}%
\usepackage{manyfoot}%
\usepackage{booktabs}%
\usepackage{algorithm}%
\usepackage{algorithmicx}%
\usepackage{algpseudocode}%
\usepackage{listings}%
\usepackage[utf8]{inputenc}
\usepackage{subcaption}
\usepackage{braket}
\usepackage{xspace}
\def\BibTeX{{\rm B\kern-.05em{\sc i\kern-.025em b}\kern-.08em
    T\kern-.1667em\lower.7ex\hbox{E}\kern-.125emX}}

\usepackage{graphicx}
\graphicspath{ {./images/} }

\newcommand{\mus}{$\mu s$\xspace}

\newcommand{\appendixref}[1]{Appendix~\ref{#1}}

\newcommand{\Thahn}{T_2^\mathrm{Hahn}}

\newcommand{\Fro}{F_{\mathrm{R/O}}}
\newcommand{\Fsq}{\mathrm{F}_{\mathrm{1Q}}}

\begin{document}

\title[Article Title]{Evaluating Three Levels of Quantum Metrics on Quantum-Inspire Hardware}


\author*[1]{\fnm{Ward} \sur{van der Schoot}}\email{ward.vanderschoot@tno.nl}

\author[1]{\fnm{Robert} \sur{Wezeman}}

\author[1]{\fnm{Pieter Thijs} \sur{Eendebak}}

\author[1,2]{\fnm{Niels M. P.} \sur{Neumann}}

\author[1,3]{\fnm{Frank} \sur{Phillipson}}

\affil*[1]{\orgname{The Netherlands Organisation for Applied Scientific Research}}

\affil[2]{\orgname{Qutech}, \orgaddress{\country{The Netherlands}}}

\affil[3]{\orgname{Maastricht University}, \orgaddress{\country{The Netherlands}}}


\abstract{With the rise of quantum computing, many quantum devices have been developed and many more devices are being developed as we speak. This begs the question of which device excels at which tasks and how to compare these different quantum devices with one another. The answer is given by \textit{quantum metrics}, of which many exist today already.
Different metrics focus on different aspects of (quantum) devices and choosing the right metric to benchmark one device against another is a difficult choice. 
In this paper we aim to give an overview of this zoo of metrics by grouping established metrics in three levels: component level, system level and application level. With this characterisation we also mention what the merits and uses are for each of the different levels.
In addition, we evaluate these metrics on the Starmon-5 device of Quantum-Inspire through the cloud access, giving the most complete benchmark of a quantum device from an user experience to date.}




\maketitle

\section*{Acknowledgements}
This project has received funding from the European Union’s Horizon 2020 research and innovation programme under grant agreement No 951852.

\section{Introduction}
Since the first experimental realisation of a 2-qubit quantum chip in 1998~\cite{Chuang:1998}, quantum computing has seen rapid developments. Nowadays, an increasing number of parties is investing in quantum computers, each with a different focus: from building them to developing applications and programming the algorithms for them. 
Over the years, the technology has evolved and current state-of-the-art gate-based and quantum annealing hardware offer over four hundred and five thousand qubits~\cite{IBM,IBM_Osprey:2022, D-Wave_5000Q}, respectfully. Roadmaps present a path to scale the number of qubits in the coming years to tens of thousands and more. The evolution of the number of qubits in the past together with the projected number of qubits shows a similar exponential scaling as Moore's law predicted for the number of transistors on integrated circuits~\cite{Moore:1965}.

As the number of qubits of quantum computers increases, so do the capabilities to run quantum algorithms for larger problem instances. However, the number of qubits is not the only metric to measure the power of quantum computers. The quality of the qubits and what type of operations we can run on them matter as well. 

Some metrics quantify individual elements of a quantum computer, such as how good single qubit operations are.
These metrics are often specific to a computational paradigm such as gate-based quantum computing. 
They are most used to quantify the performance of single components of a quantum device, and to gauge the strengths and weaknesses of a device.
Currently, no consensus exists on what the best metric or group of metrics is to quantify the power of quantum devices via the individual elements. 
This complicates comparing different quantum devices with each other. 

Other metrics describe a quantum system as a holistic unit instead of as a group of individual qubits. 
These metrics manage to measure how the different components of a system perform as a whole.
They hence test how well the device performs on running some random algorithm, however they might fail to describe the device's performance on practical applications. 
Finally, some metrics instead measure how capable a system is in solving real-world problems or running standard quantum algorithms. 
For end users of quantum device, these metrics carry the most valuable information.

Completeness, comprehensiveness and applicability of a metric are competing forces. 
Increasing one aspect often diminishes another. 
This holds especially given the wide range of available quantum devices, and the diverse technologies the quantum devices are based on. It is often the case that quantum metrics favour one hardware technology over the other, whilst they perform similarly on practical applications.

This work’s main contribution is defining three levels of quantum metrics to evaluate quantum devices with. The main user group of the three levels also differs: from quantum hardware developers, via scientists, to end-users. Previous work on quantum metrics often only focused on the lower-level metrics and then evaluated such a metric on a single device with direct low-level access to the device~\cite{quantum_volume_64:2020}, or evaluated a single metric on different backends\cite{Schoot2022EvaluatingQscore}.Our focus on three levels of metrics however gives a better overview of the performance of a quantum device for different target audiences. As our focus is on defining the three levels and showing how they highlight different aspects of a quantum device, we only evaluate them for a single device. Furthermore, we restrict ourselves to API access of the device for the higher levels, to test the performance experienced by end-users. To the best of our knowledge, we are the first to make such an extensive evaluation of a publicly accessible quantum device.

The three levels of metrics for quantum devices are component-level, system-level and application-level metrics. With these grouping of metrics into levels, we aim to standardise the current landscape of quantum metrics. We will evaluate examples of metrics in each of the three levels on the Starmon-5 device of Quantum-Inspire\cite{QI, QI:starmon5}.

This work will be structured as follows. In the next section, we discuss these three levels of quantum metrics and mention the most well-known examples of each level. In Section~\ref{sec:results} we briefly introduce the Starmon-5 device and evaluate the discussed metrics on this device. We finish this paper with a discussion of the results and an outlook.

\section{Different levels of quantum metrics}
In this section we aim to standardise the landscape of quantum metrics by dividing them up into three levels. As mentioned above, the three levels of quantum metrics are: component-level, system-level and application-level metrics. Whilst all metrics allow the benchmarking of a quantum device, each level is suitable for evaluating a device in a different way. In particular, each level is relevant for different groups of people working with quantum computing. Below we will discuss the differences between the different levels, we will indicate what kind of benchmarking each level is most suitable for, and we will give examples of well-known metrics for each level.

\subsection{Component-level metrics}
Component-level metrics tell something about small components of a quantum device. Component-level metrics are useful when one wants to understand the low-level workings of a quantum device. They are most suitable to detect strengths and weaknesses of a quantum device and they hence form vital information in the development of quantum devices. For this reason, component-level metrics are most relevant for hardware providers or low-level quantum developers.

A quantum device is made up of various components, because of which there are many different component-level metrics. The quality, number and connectivity of qubits, as well as the quality of preparation, operations and measurement are common components studied by component-level metrics. Other properties are defined by the design of the quantum system, such as the number of qubits, the qubit connectivity, and the available primitive gates. Although these properties impact the performance of the system, they are harder to capture in a single value. The following is a list of often-used component-level metrics which can be captured by a single value:

\begin{itemize}
    \item \textbf{Single-qubit gate fidelity} This single-qubit gate fidelity gives for each qubit the average error per gate applied to it. This is typically measured with randomised benchmarking~\cite{RB} or gate set tomography~\cite{Nielsen2021gatesettomography};
    \item\textbf{SPAM errors} The SPAM metric measures errors resulting from state-preparation and measurements. Disentangling the initialisation and measurement errors in a quantum system is often hard, but typically the measurement errors are dominant;
    \item\textbf{Crosstalk} The crosstalk metric gauges how different qubits of a quantum device influence one another in an undesirable way. This can be characterised by computing the probabilities of sudden transition between the basis states of the quantum device.
    \item\textbf{$\mathbf{T_1}$} The $T_1$ (or energy relaxation time) time measures the qubit lifetime, e.g., the typical lifetime in which a qubit decoheres from the $|1\rangle$ to the $|0 \rangle$ state;
    \item\textbf{$\mathbf{T_2^*}$} The $T_2^*$ time measures the qubit dephasing time. This measures the rate at which the phase of a qubit loses information; 
    \item\textbf{$\mathbf{T_2^\mathrm{Hahn}}$} The Hahn echo metric $T_2^\mathrm{Hahn}$ also measures the qubit dephasing time. It measures this dephasing time different than the $T_2^*$ value, resulting in a different metric.~\cite{Medford2012};
    \item\textbf{$Q$-factor} The $Q$-factor, or quality factor, measures the number of gates a quantum device can perform in a single $T_2^*$ time, averaged over all the qubits. This loosely measures the number of gates a quantum device can perform before a qubit is rendered useless.
\end{itemize}

Other component-level metrics are given by the two-qubit gate fidelity, the native gate set and the gate speed. We will not measure these metrics in this work. The two-qubit gate fidelity, similar to the single-qubit gate fidelity, gives the average error per two-qubit gate. Typically, this metric is only evaluated for adjacent qubits. 
The native gate set is the set of gates which can be implemented natively on a quantum device. This affects how easy it is to implement arbitrary operations. 
If the gate-set is universal, we can in principle implement any operation~\cite[Section~4.5]{Nielsen2000}.
However, some universal gate sets allow for a more efficient decomposition compared to others.
The gate speed describes the time it takes to implement a quantum gate.
Due to decoherence of qubits, this time interval in which we can apply gates is limited. With higher gate speeds, we can implement more gates in the same time interval, resulting in more applied gates within the decoherence lifetime of the qubits.

Improving a device's performance on one component-level metric can affect its performance on another component-level metric, as becomes apparent when comparing the gate speed and single-qubit gate fidelity: we can increase the gate fidelity by taking more time to perform the gate, but the increased duration will directly reduce the gate speed. Various error-mitigation techniques allow improvements of the performance of the system~\cite{BMZKT_ProbabilisticErrorCancellation:2022,GHLMZ_ZeroNoiseExtrapolation:2020}.
An example is readout error-mitigation~\cite{measurement_error_mitigation}, which can greatly improve the performance of various applications but requires more work from the user-side.

\subsection{System-level metrics}
System-level metrics describe a system as a holistic unit. They measure how the quantum device performs at solving simple quantum algorithms. These metrics are most suitable to determine whether the different components of a quantum device collaborate as expected and give the expected results. Because of this, these metrics are most relevant for hardware providers, quantum developers or low-level end users.

Below we consider three different quantum metrics which measure respectfully the speed and quality of a quantum system. 
Two of them require the device to execute a random algorithm and performance is measured by comparing the results to the expected outcomes. 
The other metric considers the system stability over time. 

\begin{figure}
    \centering
    \includegraphics[width=0.8\columnwidth]{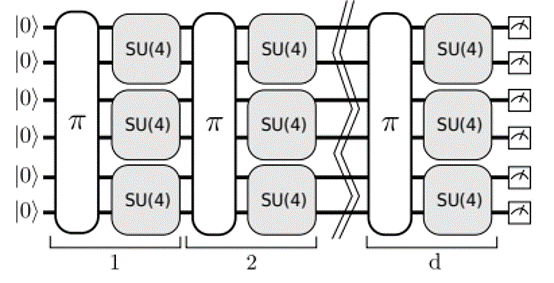}
    \caption{A schematic overview of the quantum volume circuit. Each layer consists of a random permutation and parallel two-qubit random unitaries \cite{quantumvolume}.}
    \label{fig:quantum_volume}
\end{figure}

The first considered metric is the \textbf{quantum volume}~\cite{quantumvolume}, which measures the quality of performance of a quantum device. Specifically, it measures the largest square circuit a quantum processor can successfully execute on average. This is gauged by running multiple random circuits consisting of layers of random two-qubit operations applied on random pairs of qubits.
Figure~\ref{fig:quantum_volume} gives a graphical representation of the random circuit used in the quantum volume metric. 
The operation indicated with $\pi$ represents some permutation so that the right qubits are adjacent for the next layer of two-qubit gates. The $SU(4)$ blocks indicate random $2$-qubit unitaries. The quantum volume is defined as $2^d$, where $d$ is largest such that a random square circuit of $d$ qubits and $d$ consecutive layers is successfully run on the quantum device. In this definition, a successful run indicates that the \textit{heavy-output generation problem} is correctly solved, which is the problem of sampling outputs with a higher than median probability. 
Solving the heavy-output generation problem requires knowledge of the outcome probabilities and hence requires classical simulations. Because of this, the quantum volume metric is not scalable to large problem instances, which limits its applicability.


While the quantum volume metric gives an indication of the performance of a device, it discards the time-aspect of the computation. 
The second system-level metric, the \textbf{system stability} metric, takes the time aspect implicitly into account. 
Quantum systems are usually constructed and tuned to a certain level of performance. 
After the tuning, the performance will degrade over time, either slowly (e.g.,~drift of parameters) or suddenly (e.g.,~a charge jump).
No well-defined metric for system stability exists yet, so we analyse it by looking at the $T_2^*$ value over time. We then define the system stability metric as how constant the $T_2^*$ value is over time. 

\begin{figure*}
    \centering
    \includegraphics[width=\textwidth]{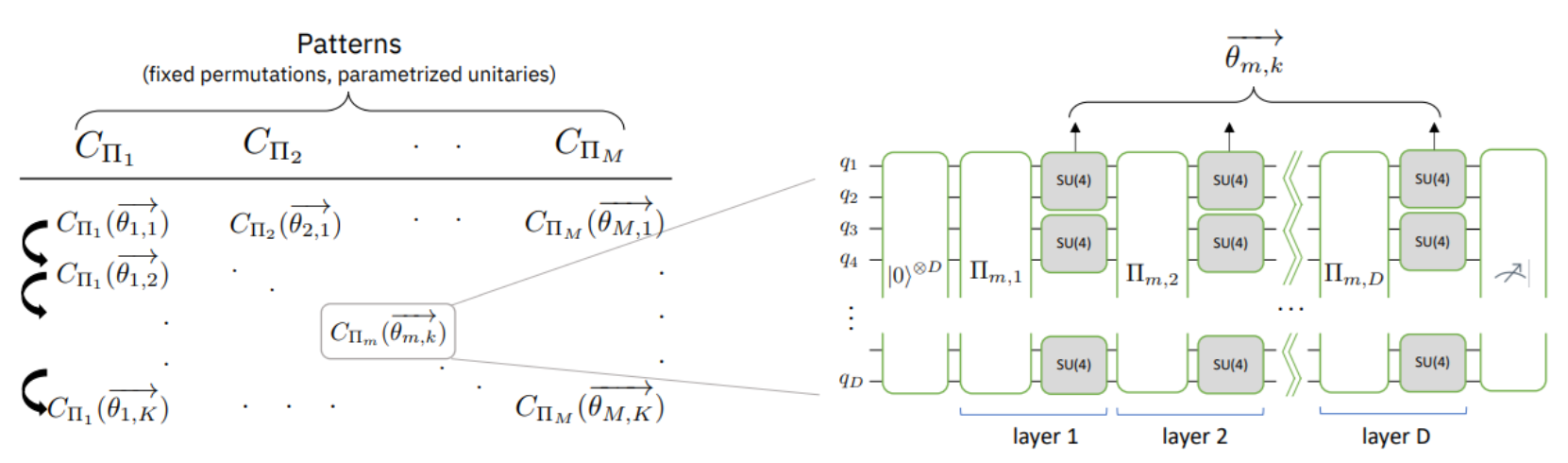}
    \caption{A schematic overview of the Circuit Layer Operations Per Second (CLOPS) system-level metric. We consider $M$ parameterised quantum volume circuits in parallel and $K$ sequential executions. The parameters $\theta_{m,k}$ depend on the output of the $k-1$-th sequential circuit of the $m$-th parallel instance \cite{CLOPS}.}
    \label{fig:clops}
\end{figure*}

The \textbf{Circuit Layer Operations Per Second (CLOPS)}~\cite{CLOPS} metric is another system-level metric that explicitly measures the speed with which a quantum device can perform quantum operations.
Specifically, it measures the number of quantum volume layers a quantum device can execute per second. It does this by running many different quantum volume circuits and measuring the total execution time. In this way it measures the speed of a quantum device when it is operating at its highest quality performance in terms of the quantum volume. In 
In the computation of the CLOPS metric, the application of a real-world algorithm is mimicked as well as possible. In particular, aspects as circuits depending on earlier measurements and parallelism are taken into account. This is done by having parameterised quantum volume circuits which depend on one another - the outcome of one quantum volume circuit determines the parameters of the next - and parallel circuits, each of which is run with separate parameter updates, respectively.
The time spent on data transfers between classical CPU, cache and main memory, and feedback of classical intermediate computations back into the quantum circuit, are part of the execution time of real-world algorithms and are hence also taken into account by the CLOPS metric. 
Figure~\ref{fig:clops} gives a schematic overview of the CLOPS circuit.
The CLOPS system-level metric is given by 
\begin{equation}
    CLOPS = \frac{M\times K\times S\times D}{T_\mathrm{total}},
\end{equation}
where $M$ is the number of parallel circuits, $K$ the number of parameter updates, $S$ the number of shots per circuit, $D$ the number of layers/qubits in each quantum volume circuit
and $T_\mathrm{total}$ the total calculation time. 
Note that the CLOPS system-level metric also requires classical simulations to compute the quantum volume and hence faces similar limits as the quantum volume system-level metric.

\subsection{Application-level metrics}
Application-level metrics describe the performance of a quantum device on specific problems or algorithms.  
The focus of application-level metrics is on the actual use in practice.
These metrics can hence be used to determine which device is most suitable for a certain application.
Because of this, application-level metrics are most relevant for end users and investors.
In this work, two different application-level metrics are considered.

The first is the \textbf{QED-C} metric~\cite{Lubinski:2021}. 
This metric considers various tutorial algorithms, quantum subroutines and end-user applications to measure the performance of a device. 
Eleven algorithms that are at the heart of many quantum algorithms are considered, of which examples include the Deutsch-Jozsa algorithm~\cite{DeutschJozsa:1992}, the quantum Fourier transform~\cite{Coppersmith:1994} and Hamiltonian simulation~\cite{Feynman1982,Lanyon2010}. 
The results are shown in a volumetric graph, with colours indicating the fidelity, and the axes showing the depth and width of the respective circuits.

The second application-level metric is the \textbf{Q-score}~\cite{Qscore}. 
This metric evaluates how well a system performs on the Max-Cut problem, specifically on random Erd{\"o}s-R{\'e}nyi $(N,\frac{1}{2})$-graphs. 
The outcome is a single value corresponding to the largest graph size for which the found cut is significantly larger than one found by a random approach. 
As classical devices might outperform quantum devices, given sufficient time, we slightly modify the Q-score metric by adding a time constraint of 60 seconds for the total computation.
Earlier work already benchmarked different quantum annealing devices against classical and hybrid solutions using this time constraint~\cite{Schoot2022EvaluatingQscore}. 
An extra advantage of the Q-score is that it can be used to benchmark different paradigms of quantum computation, as well as classical solution methods.
Recently, the Q-score has been extended to consider other problems as well, such as the Max-Clique problem~\cite{QscoreMaxClique}.

\section{Results}
\label{sec:results}


The Quantum-Inspire system~\cite{Last2020} is an online cloud-based quantum platform. 
On the launch in April 2020 a classical simulator and two hardware backends where available: Spin-2 and Starmon-5. 
At the time of writing only the Starmon-5 backend is publicly available through the online API in the cloud.

The Starmon-5 device is a five-qubit device with in total four qubit connections. The centre qubit (qubit 2) of the device is connected to the other four qubits via the two-qubit CZ gate.
The device itself evolves slowly over time. Due to this drift and due to retuning or upgrades of the device, metric values can change over time. The system stability is discussed in Section~\ref{section:system level metrics}.
The results in this work were obtained in the months November and December of 2022. 

Below we present the result of benchmarking the above metrics on this device. In this work, we want to obtain metric results in a way which is representative for the sort of users of the respective metric. As mentioned above, component-level metrics are mostly interesting to hardware providers and low-level quantum developers. These groups often interact with quantum devices through direct access to the device their working with, because of which we calculated the component-level metrics using direct access to the device. System- and application-level metrics are more of interest to quantum algorithm developers and end users. These groups usually interact with quantum devices through cloud access. That is why we have computed the system- and application-level metrics to the API cloud access of the Starmon-5 device.

It should be noted that the Starmon-5 device is a rather small device. Because of this, it is expected that the device will output relatively low and uninteresting scores on the system- and application-level metrics. Still, we believe it is important to show the evaluation of the Starmon-5 device using these metrics. The main reason for this is to show that such an extensive evaluation of a single device is indeed possible using all these metrics.

\subsection{Component-level metrics}
\label{sec:component_level_results}
In this section we present an evaluation of the component-level metrics on the Starmon-5 system.
Specifically, the single-qubit gate fidelity, the SPAM error rates, crosstalk, the coherence times of each of the qubits and the $Q$-factor of the Starmon-5 system are shown. The settings for each experiment can be found in \appendixref{section:experiment_settings}.

\begin{figure}
  \centering
    \includegraphics[width=0.8\columnwidth]{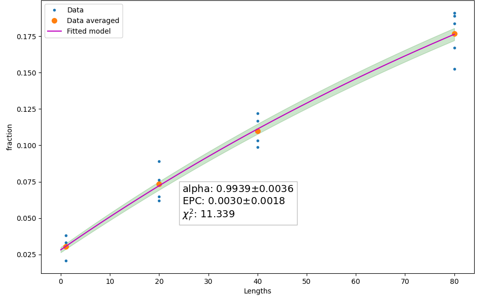}
  \caption{The fraction of incorrect measurements per Clifford sequence length in the randomised benchmarking protocol for qubit~$1$. The fitted model is $F(N)=A\alpha^N+B$.}
  \label{figure:rb_fit}
\end{figure}

The single-qubit fidelities of the Starmon-5 system were measured using standard randomised benchmarking with Clifford gates~\cite{RB}. This method works by applying sequences of random single-qubit gates which together should yield the identity. For details, please see \appendixref{section:rb}.
In our experiments, we considered Clifford sequences of length $N=1, 20, 40, 80, 120$, with 10 sequences per length.
Figure~\ref{figure:rb_fit} shows the obtained results for qubit~$1$. In this diagram, the horizontal axis depicts the length of the Clifford sequences, while the vertical axis denotes the fraction of incorrect measurements. With this, the model $F(N)$ as defined in \appendixref{section:rb} can be fitted. From this model, the growth factor $\alpha$ and the corresponding single-qubit fidelity can be computed. The single-qubit fidelity for each qubit can be found in Table~\ref{table:component_metrics}.

\begin{table}
\caption{Overview of single-qubit system metrics. The $\Fsq$ is the single-qubit gate fidelity for the native gates and $\Fro$ the readout fidelity.}
\centering
\begin{tabular}{|l|c|c|c|c|c|}
     \hline
     Qubit & $T_1$ $(\mu s)$ & $T_2^*$  $(\mu s)$ & $T_2^\mathrm{Hahn}$  $(\mu s)$&  $\Fsq$ (\%) & $\Fro$ (\%)\\ [.5ex]
     \hline\hline
0 NW & $15.45 \pm 0.25$ & $13.29 \pm 0.62$ & $28.39 \pm 0.64$& $99.798 \pm 0.040$& 96.7  \\ 
1 NE    & $15.95 \pm 0.54$ & $24.68 \pm 0.81$ & $34.50 \pm 1.10$&$99.827\pm 0.040 $& 96.8  \\ 
2 C   & $19.42 \pm 0.40$ & $21.40 \pm 1.00$ & $31.30 \pm 0.38$& $99.812 \pm 0.036 $& 97.5  \\ 
3 SW   & $22.74 \pm 0.82$ & $21.40 \pm 1.90$ & $43.90 \pm 1.70$& $99.828 \pm 0.034 $& 98.4  \\ 
4 SE  & $12.21 \pm 0.26$ & $16.20 \pm 2.30$ & $24.15 \pm 0.51$& $99.868 \pm 0.071$& 96.4  \\ 
\hline
\end{tabular}
\label{table:component_metrics}

\end{table}

\begin{figure}
  \centering
    \includegraphics[width=\columnwidth]{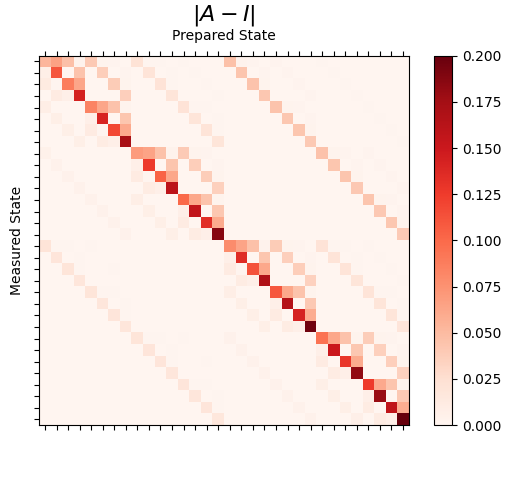}
  \caption{Readout assignment matrix for the Starmon-5 device. Each circuit was measured with $16{,}384$ shots. Displayed is the absolute value of the difference of the assignment matrix and the ideal diagonal assignment matrix.}
  \label{figure:starmon5_readout_assignment_matrix}
\end{figure}

For transmon systems, the SPAM errors are typically dominated by the measurement errors, as the measurement errors are usually much larger than the initialisation errors. However, our approach cannot distinguish between both errors. Hence, we present a single number for the combined error, which we call the readout error.
To measure the readout errors on Starmon-5, two experiments are run, in which the system is prepared in the states $|00000\rangle$ and $|11111 \rangle$ respectively, after which the resulting state is measured.
This measurement  yields the $2 \times 2$ readout assignment matrix $M$ for each qubit, in which the entry $M_{ij}$ denotes the probability of measuring state $|j\rangle$ when the qubit was prepared in state $|i\rangle$. From this readout assignment matrix the readout fidelity $\Fro$ can be computed as
\begin{align}
\Fro = 1 - (M_{10}+M_{01})/2 .
\label{equation:Freadout}
\end{align}
The result of the measurement is in Figure~\ref{figure:starmon5_readout_assignment_matrix}, and the calculated fidelities are in Table~\ref{table:component_metrics}.
The measured readout fidelities are comparable to the numbers reported on the Quantum Inspire website~\cite{QuantumInspire}. 
It should be noted that repeating readout fidelity measurements on the qubits results in a higher variance in the results than expected for a stable system.

To characterise the crosstalk between the qubits, one experiment for each of the $2^5$ basis states has to be run. In each experiment, the system is prepared in the respective basis state, after which the resulting state is measured. From this, the probability of transitioning from one basis state to the other can be computed. This results in the full $2^5 \times 2^5$ readout assignment matrix, as depicted in Figure~\ref{figure:starmon5_readout_assignment_matrix}. Each of the $2^5$ basis states are denoted on both axes. Note that in an ideal quantum device, no crosstalk should take place, resulting in a completely diagonal readout assignment matrix.
%








%
%

\begin{figure}
  \centering
    \includegraphics[width=\columnwidth]{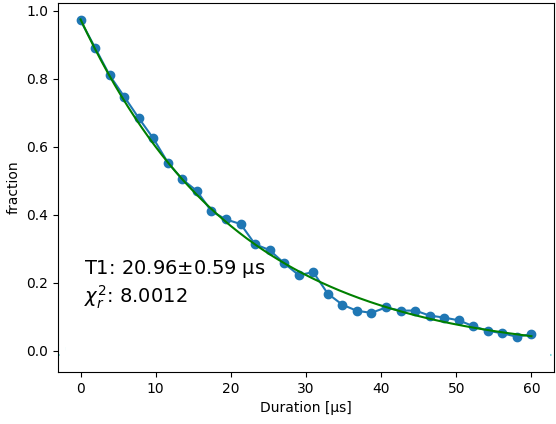}
  \caption{$T_1$ measurement for qubit 2 of the Starmon-5 system. The fitted model is $F(t)=A +B e^{-t/T_1}.$}
  \label{figure:T1_measurement}
\end{figure}

\begin{figure}
  \centering
    \includegraphics[width=\columnwidth]{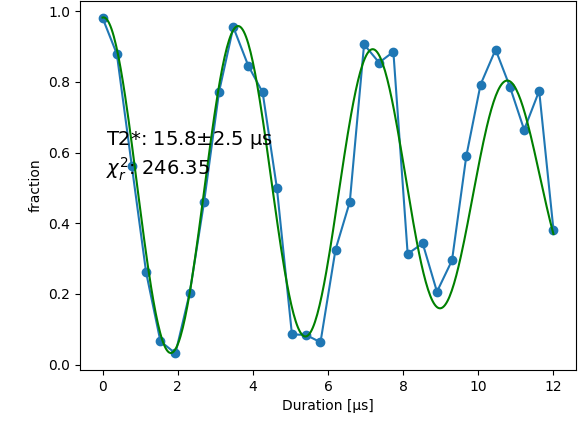}
  \caption{$T_2^*$ measurement for qubit 3 the Starmon-5 system. The fitted model is $F(t) = A + B e^{-t/T_2^*} \sin(\omega t + \phi)$.}
  \label{figure:T2star_measurement}
\end{figure}

\begin{figure}
  \centering
    \includegraphics[width=\columnwidth]{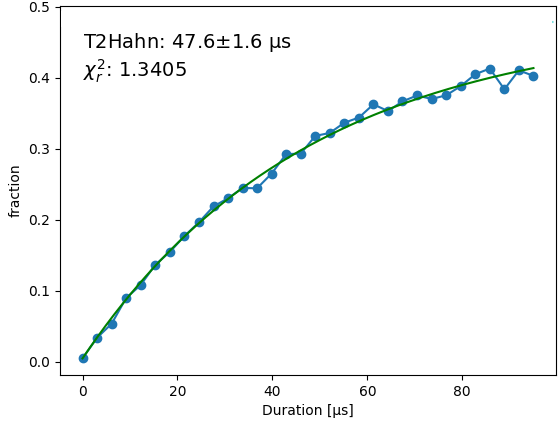}
  \caption{$\Thahn$ measurement for qubit 3 the Starmon-5 system. The fitted model is $F(t)=A +B e^{-t/T_2^{\mathrm{Hahn}}}.$}
  \label{figure:T2_Hahn_measurement}
\end{figure}

The coherence times of the qubits were computed by using standard coherence time experiments. In each experiment, the number of $|1\rangle$ measurements of a certain quantum circuit are computed over time. This number is then fitted to a model, which yields the required coherence time. Depending on the used quantum circuit, a different coherence time is obtained. More details can be found in \appendixref{section:coherence_experiments_details}. The coherence times for each of the qubits of the Starmon-5 device can be found in Table~\ref{table:component_metrics}. As is characteristic for transmon qubits, the $T_1$ times are not much larger than the $T_2^*$ times, while adhering to the relation $T_2^*\le 2T_1$~\cite{Traficante1991,Levitt2008}. The values for the coherence times $T_1$, $T_2^*$ and $T_2^\mathrm{Hahn}$ for the five qubits show quite some variation. 
This can be the result of the non-uniformity of the device or the tuning of the individual qubits. Because of the variation in the performance of individual qubits, the performance of executing a quantum algorithm will depend on the mapping of the algorithmic qubits onto the physical qubits.
For completeness, figures~\ref{figure:T1_measurement}, \ref{figure:T2star_measurement} and \ref{figure:T2_Hahn_measurement} show the $T_1$, $T_2^*$ and $\Thahn$ experiments for qubit 2, 3 and 3 respectively.
For the $T_2^*$ experiments, there are quite some outlier datapoints for larger waiting durations. 
This will be discussed in more detail in the system stability metric in Section~\ref{section:system level metrics}. 
The fitted model and $T_2^*$ value do not seem to be influenced strongly by the outlier datapoints.

A native single qubit rotation on Starmon-5 takes 20~ns~\cite{Starmon5Factsheet}. 
The $Q$ factor for the Starmon-5 system, using the average $T_2^*$ time of $19.30$ all five qubits, can hence easily be computed as $970$. 

Table~\ref{table:component_metrics} reports the single-qubit component-level metrics for Starmon-5. 
Details on how these values were obtained are in Appendix~\ref{section:coherence_experiments_details}.

\subsection{System-level metrics}
\label{section:system level metrics}
We have computed the quantum volume and the CLOPS on the Starmon-5 device. 
We also analysed the system stability by analysing the $T_2^*$ values over time. 
For the CLOPS metric, we only incorporated the quantum execution time and the classical pre- and post-processing time.
We omitted the queuing time and the time the system was offline as this would skew the results. 

For the Quantum Volume metric, we used $100$ different circuits for each depth $d$, and we executed each circuit $100$ times.
For the CLOPS metric we used $100$ different parameterised circuits, each run with $10$ parameter updates. Again, each parameter update was executed $100$ times. 
We used the built-in Aer-simulator from Qiskit~\cite{aer} to do the classical simulations for the quantum volume metric.

We obtained a quantum volume of $4$, corresponding to $2$ layers of random two-qubit gates and a CLOPS of 372 circuit layer operations per second. The quantum volume of 4 indicates that only really small circuits with at most two consecutive 2-qubit gates on two different qubits can be applied reliably on the Starmon-5 device. In contrast, the CLOPS value indicates can perform such circuits in very short time frames under $1/100$ of a second.

In addition, we tested the system stability of the Starmon-5 device. 
The Starmon-5 device, as any transmon-device, has a drift in the operating parameters. 
For this reason the system parameters are re-tuned periodically. 
The drift results in a small variation in the performance metrics over time. 
In addition, on shorter timescales, the system can have larger instabilities. 

\begin{figure*}[ht!]
  \centering
    \includegraphics[width=\textwidth]{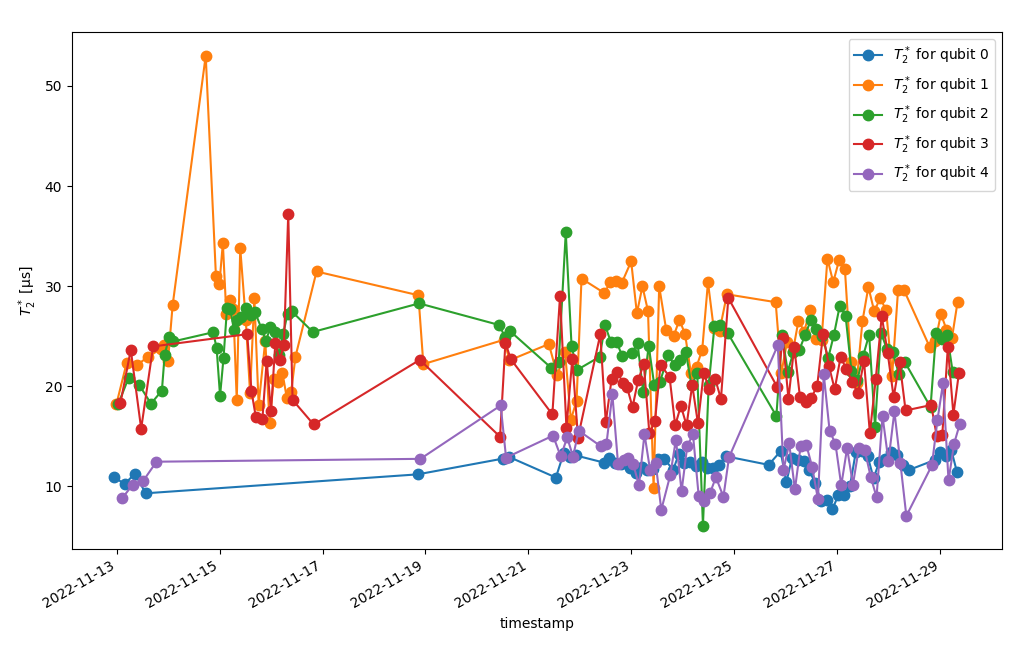}
  \caption{$T_2^*$ values of the Starmon-5 device over time.}
  \label{figure:t2star_vs_time}
\end{figure*}

Figure~\ref{figure:t2star_vs_time} shows the variation in the $T_2^*$ value over time, in which significant variations can be found. Seeing that each data point was obtained with $4,096$ shots, the impact of quantum randomness should be minimal. It can hence be concluded that the Starmon-5 device is a rather unstable device.



\subsection{Application-level metrics}
\begin{figure}[ht!]
  \centering
    \includegraphics[width=\columnwidth]{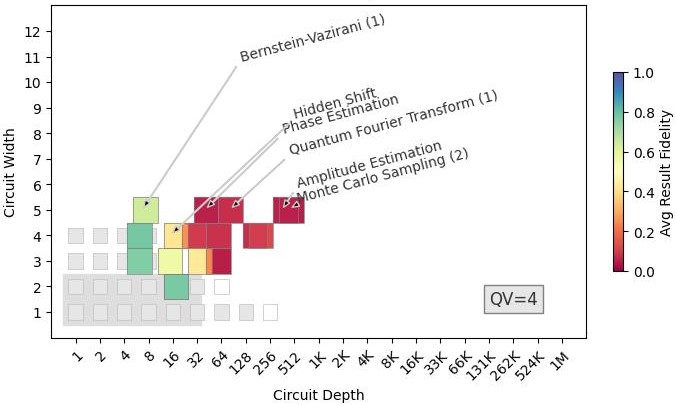}
  \includegraphics[width=\columnwidth]{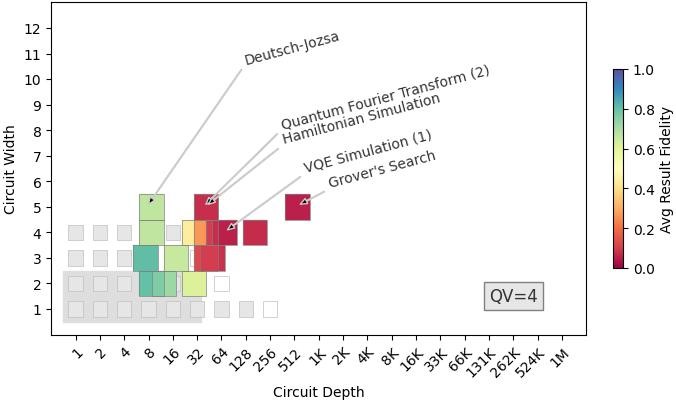}
  \caption{The QED-C metrics computed on the Starmon-5 device. The top and bottom figure show two different sets of algorithms.}
  \label{figure:QED-C}
\end{figure}

We have computed the QED-C metric and Q-score on the Starmon-5 device. 
To compute the QED-C metric, we used the source code provided by the QED Consortium~\cite{QED-C_source} for $2$ to $5$ qubits. 
Figure~\ref{figure:QED-C} shows the results. 
For each of the eleven algorithms, circuits at varying widths and depths are run. Each square in the diagram indicates an experiment for a certain algorithm, with depth and width indicated by the axes. The colour of the square indicates the obtained fidelity for the experiment. 
The grey blocks indicate the performance of the device on the quantum volume metric, with the possible results being successful (grey) or unsuccessful (white). 

From these results it can be seen that the Starmon-5 device solves the tutorial algorithms such as the Bernstein-Vazirani algorithm and the Deutsch-Jozsa algorithm with high fidelities, even when using all qubits. 
We do see that the device achieves a low fidelity on the quantum subroutines such as the quantum Fourier transform and amplitude estimation. 
We see a similar performance for the end-user applications such as VQE simulation and Grover's search algorithm. 

For the Q-score, we used the source code provided by ATOS~\cite{ATOS_source}.
We use the quantum approximate optimization algorithm (QAOA)~\cite{QAOA:2014} to solve the Max-Cut problem. 
We consider Erd{\"o}s-R{\'e}nyi $(N,\frac{1}{2})$-graphs with $N\in\{2,3,4,5\}$ nodes, corresponding to the number of used qubits. 
For each graph size$N$, we consider $5$ different graphs, average the results and compare those to a random approach. 
We found a Q-score of $2$, which was largely due to the time constraint of 60 seconds. If we were to release this constraint, we end up with the maximal Q-score value of $5$ with corresponding time duration of over an hour. This maximal value is not a surprise given the limited amount of possibilities for such small graph sizes. As QAOA works by aptly trying out different possibilities depending on earlier outcomes, the result without time constraint has no actual meaning: with a limited number of possibilities, QAOA becomes a smart brute-force search algorithm, which will always give the correct solution given sufficient time.

\section{Discussion}
In this work, we identified three levels of quantum metrics and evaluated metrics from each level on the Starmon-5 quantum device of Quantum Inspire. 
The first level consists of component-level metrics that quantify the individual elements of a device, such as qubit decoherence times and single and two-qubit gate fidelities. 
The second level consists of system-level metrics that describe a device as a whole, for instance in terms of the system stability or by quantifying how well the device solves some (randomised) algorithm. 
The third and last level consists of application-level metrics that describe the power of a quantum device in running standard quantum algorithms or solving problems. 

Our evaluation of multiple metrics on the Starmon-5 device presents a good overview of its capabilities. 
We evaluated the component-level metrics using direct access to the device and the system-level and application-level metrics via the online API.
This is in line with how end-users would use the device. Component-level metrics are most useful in settings with direct access to the device, while system-level and application-level metrics are more relevant for users through the API access.
It should be noted that the use of the API access has its limitations. In particular, queue times and regular re-calibrations prevented us from running multiple instance of some algorithms for longer periods of time.
It is expected the results for the system- and application-level metrics might turn out higher if they were evaluated using direct access and tuning of the implementation of the circuits. 
 
The found values for the component-level metrics are in accordance with what we expect for transmon devices. 
We found high fidelities for single qubit gates and slightly lower fidelities for readout. The most limiting components of the device seem to be its decoherence times, which were in the orders of microseconds. Comparing this to the CLOPS metric, one can see that the decoherence times are much lower than the time required to apply gates. Because of this, the decoherence time limit the applicability of the device.

The system- and application-level metrics indeed confirmed that the applicability of the device is rather minimal. In particular, a result of 4 was found for the quantum volume metric indicating that circuits of up to $2$ computational layers with $2$ qubits can be faithfully executed. 
For reference, this value is lower than found on other 5-qubit devices, such as those operated by IBM~\cite{QV:IBM}, though these higher values were achieved with direct access to the device and high levels of optimisation in the operations. This observation is confirmed by the QED-C metrics, which show that only algorithms with low depths and widths can be applied reliably.

From all results together, we can see that the Starmon-5 is currently limited in its performance to run quantum circuits. This is not a surprise, as Starmon-5 is quite a small device and it is the first superconducting device designed by QuTech. The Starmon-5 device is currently mainly interesting for research parties, who wish to experiment with specific qubits and gates, or small toy circuits.

Our work has made an attempt at structuring the zoo of quantum metrics which currently exist in literature. Many of these metrics aim to become \textit{the} quantum metric which should be used as the one and only metric in the development of quantum devices. We do not believe in the existence of such a single quantum metric, and actually see the strength in having multiple quantum metrics, each with a different purpose. Specifically, all three levels of quantum metrics are very relevant for the development of quantum computing, albeit each level for a different group of developers. Even within a single level, multiple metrics are relevant for benchmarking a quantum device.

In this work, we have evaluated a single quantum device using a multi-benchmark approach. Using this approach, we have obtained a full description of the device with its strengths, weaknesses and potential applicability. While such an evaluation is very interesting, it should be emphasised that for many purposes this is not the optimal way of benchmarking a device. Oftentimes, a single or a few metrics are more suitable for benchmarking the device. As an example, a couple of application-level metrics are more than sufficient for end users, while for hardware providers a list of component-level metrics suffices.

For further research, it would be interesting to repeat our analysis on different devices to see how the Starmon-5 compares to other publicly available quantum devices. The above experiments, together with the relevant citations, should yield sufficient explanation of how such an evaluation could be run. In this evaluation, it would be interesting to consider both devices that use superconducting qubits, as well as devices that use different hardware technologies. The first would show how well Starmon-5 manages to utilise the strengths of superconducting qubits, while the second would show how superconducting devices compare to devices from other hardware technologies. In particular, it will be interesting to see how much better Starmon-17, the successor of Starmon-5, will perform. 

\section*{Data Availability}
All data generated or analysed during this study are included in this published article (and its supplementary information files).

\appendix

\section{Component-level metrics details}
\label{section:component measurements}
In this appendix we discuss the specific experiments used for the component-level metrics. All experiments are performed by executing a quantum circuit while varying one or more of the circuit parameters.

\subsection{Experiment settings}
\label{section:experiment_settings}
Unless stated otherwise, each data point was run with $4,096$ shots. 
In addition, the following setting were used for each experiment:
\begin{itemize}
\item \textbf{$\mathbf{T_1}$} Number of different wait-durations: $32$, linearly spaced between 0 and the maximum wait-duration of 60 \mus;
\item \textbf{$\mathbf{T_2^*}$} Number of different wait-durations: 32, linearly spaced between 0 and the maximum wait-duration of 24 \mus, artificial oscillation frequency $0.125$ MHz;
\item \textbf{$\mathbf{T_2^\mathrm{Hahn}}$} Number of different wait-durations (total wait-duration of circuit): 32, linearly spaced between 0 and the maximum wait-duration of 120 \mus;
\item\textbf{Single-qubit RB} Lengths of Cliffors sequences: $N \in\left[1, 20, 40, 80, 120\right]$. Number of sequences per length: 10.
\end{itemize}

\subsection{Randomised benchmarking}
\label{section:rb}
For each qubit of the Starmon-5 device, the single-qubit fidelity is computed via standard randomised benchmarking. The standard randomised benchmarking protocol~\cite{Magesan2011RB} consists of circuits that contain a sequence $C_j$ of $N$ random single-qubit Clifford gates. 
For each sequence of random Clifford gates $C_j$, there exists a single-qubit Clifford gate $C_{N+1}=(\prod C_j)^{-1}$, due to the definition of Clifford gates.
For an ideal quantum system, the combined circuit with gates $C_j$, $j=1, \ldots, N+1$ acts as the identity. 
In practice, due to errors in the gates, errors will increase exponentially with the number number of Cliffords $N$, resulting in incorrect measurement results.
For each value of $N$ we generate multiple sequences of random Cliffords and apply them to the qubit initialised in state $|0\rangle$. For each sequence, the resulting qubit state is measured. 
The following model can then be fitted to the data
\begin{align}
F(N) = A \alpha^N + B,     
\end{align}
where $F(N)$ is the fraction of $|1\rangle$ measurements and $A$ and $B$ are constants to be determined. Notice that $F(N)$ exactly counts the fraction of incorrect measurements.
From the parameter $\alpha$ one can calculate the error per Clifford $r$ (EPC) using the equation
\begin{align}
r = \frac{1}{2}(1-\alpha).
\end{align}

The single-qubit gate fidelity is determined from the average number of native gates required for a random Clifford, which is $1.875$ on Starmon-5.
The single-qubit gate fidelity can hence be computed as $\Fsq=(1-r)^{(1/1.875)}$.

\subsection{Single-qubit coherence time experiments}
\label{section:coherence_experiments_details}
\begin{figure}
    \centering
 \begin{subfigure}{0.97\columnwidth}
  \centering
    \includegraphics[width=\columnwidth]{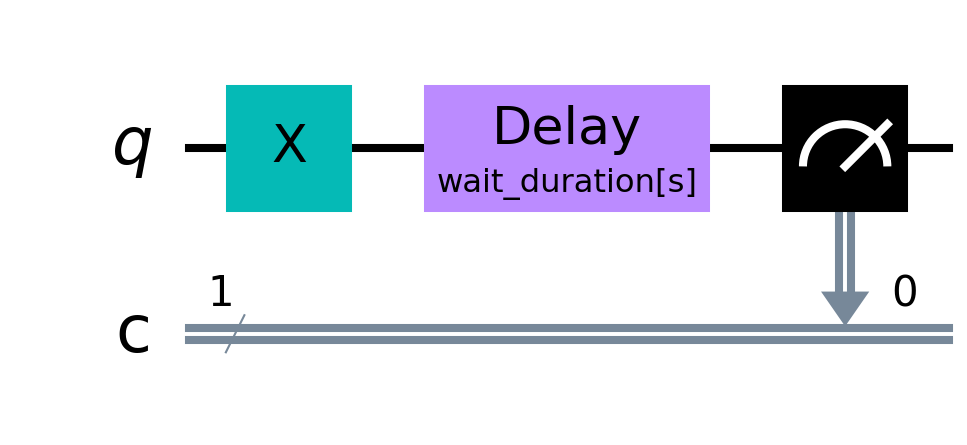}
    \caption{$T_1$ circuit}
    \end{subfigure}
 \begin{subfigure}{0.97\columnwidth}
  \centering
    \includegraphics[width=\columnwidth]{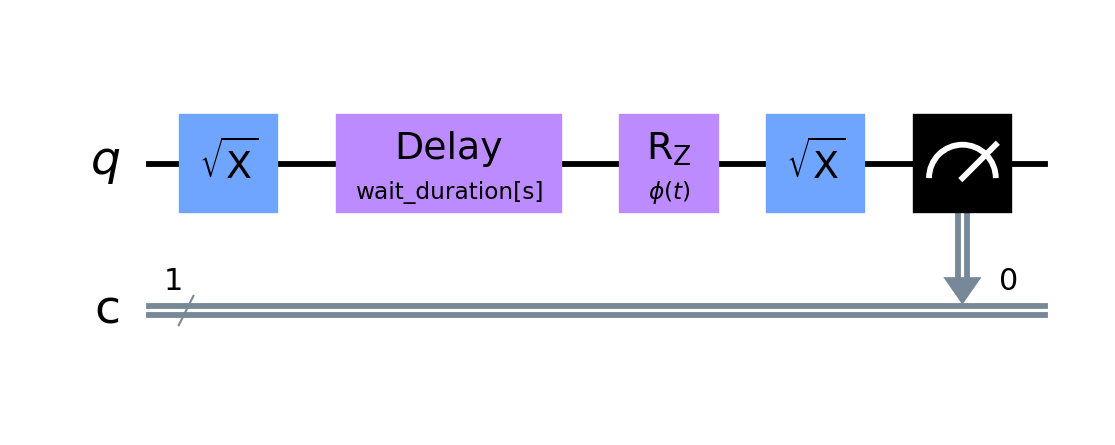}
    \caption{$T_2^*$ circuit}
\end{subfigure}
 \begin{subfigure}{0.97\columnwidth}
  \centering
    \includegraphics[width=\columnwidth]{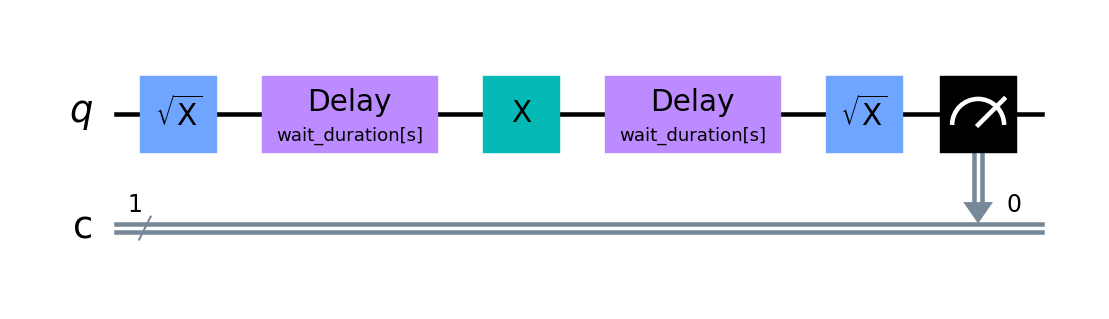}
    \caption{$T_2^\mathrm{Hahn}$ circuit}
\end{subfigure}
\caption{Quantum circuit diagrams for the single-qubit coherence measurements.}
    \label{fig:coherence circuits}
\end{figure}

In this work, three different single-qubit coherence times are computed: the $T_1$, $T_2^\mathrm{Hahn}$ and $T_2^*$ time. These values are constants used in a function fitted to the data, computed as the number of $|1\rangle$ measurements after applying a certain quantum circuit. The quantum circuit for each coherence time experiment can be found in Figure~\ref{fig:coherence circuits}.

For the $T_1$ measurement, the qubit is prepared in the $|1\rangle$ state by applying an $X$-gate to the $|0\rangle$ state. 
The qubit is then measured after a variable waiting duration, called the wait-duration. Typically, a wait-duration of up to a few times the $T_1$ time is used.
The system remains idle during the waiting stage, but the qubit can decay from the $|1\rangle$ state to the $|0\rangle$ state.
This decay results in an exponential decay of the number of measured $|1\rangle$ states. The number of measured $|1\rangle$ states for variable wait-durations $t$ is fitted to the function $F(t)=A +B e^{-t/T_1}$, from which the $T_1$ time can be computed.

To measure the $T_2^*$ time of the system, an $\sqrt{X}$-gate is applied to the $|0\rangle$-state to bring the system to an equal superposition of the $0$- and $1$-state. 
Afterwards, an $R_Z(\phi)$-rotation is applied, again after a waiting stage with variable waiting time. The $R_Z(\phi)$ is applied with an angle of $\phi(t) = \sin(\phi_R t)$ that depends on the wait-duration $t$ and the artificial frequency shift $\phi_R$. Lastly, a $\sqrt{X}$-gate is applied to map the system back. 
The result of the circuit is a damped oscillation with oscillation frequency $\phi_R$. 
The oscillation frequency is chosen such that in the total measurement a few oscillations occur so that we can properly fit the model.
The artificial frequency shift~\cite{Watson2017} is added to the circuit to prevent confounding between the damping of the signal (due to the dephasing) and a very low frequency oscillation (due to a frequency mismatch between the driving signal and the qubit resonance frequency). After performing all the experiments, the number of $|1\rangle$ states for variable wait-durations is fitted to the functions $F(t) = A + B e^{-t/T_2^*} \sin(\omega t + \phi)$, from which the $T_2^*$ time can be computed.

The $\Thahn$ time is measured in a similar fashion as the $T_2^*$ time. Again, a $\sqrt{X}$-gate is applied to the $|0\rangle$ state, after which the system remains idle for a variable waiting time. Then, an $X$ gate is applied as a single refocusing pulse. Then, the system remains idle for the same variable wait-duration, after which another $\sqrt{X}$ is finally applied. An noiseless device would always measure the resulting state as $|0\rangle$, but due to noise the number of $|1\rangle$ measurements for increasing wait-durations $t$ follows an exponential increase. By fitting this number to the function $F(t)=A +B e^{-t/\Thahn}$, the $\Thahn$ time can be computed.

\section*{Competing interests}
The research leading to these results received funding from the European Union's Horizon 2020 research and innovation programme under grant agreement No 951852.


\bibliography{journal}

\begin{thebibliography}{37}
\bibitem[\protect\citeauthoryear{Qiskit}{}]{aer}
\begin{botherref}
\oauthor{\bsnm{Qiskit}}:
{Qiskit/qiskit-aer: Aer is a high performance simulator for quantum circuits
  that includes noise models}
\end{botherref}
\endbibitem
\end{thebibliography}

\end{document}